# Laboratory experiment on Region-1 field-aligned current and its origin in low-latitude boundary layer

**I F Shaikhislamov, Yu P Zakharov, V G Posukh, E L Boyarintsev, A V Melekhov, V M Antonov and A G Ponomarenko**

Dep. of Laser Plasma, Institute of Laser Physics SB RAS, pr.Lavrentyeva 13/3, 630090, Novosibirsk

E-mail: ildars@ngs.ru

**Abstract**
In previous experiments by the authors on magnetic dipole interacting with laser-produced plasma a generation of intense field aligned current (FAC) system on Terrella poles was observed. In the present report a question of these currents origin in a low latitude boundary layer of magnetosphere is investigated. Experimental evidence of such a link was obtained by measurements of magnetic field generated by tangential drag and sheared stress. This specific azimuthal field was found to have quadruple symmetry and local maximums inside magnetosphere adjacent to boundary layer. Cases of metallic and dielectric dipole covers modeling well conductive and non-conductive ionosphere revealed that FACs presence or absence results in different amplitude and spatial structure of sheared field. Current associated with the azimuthal field flows upward at the dawnside, and toward equator plane at the duskside. It was found to coincide by direction and to correspond by amplitude to a total cross-polar current measured independently. Results suggest that compressional and Alfven waves are responsible for FAC generation. The study is most relevant to FAC generation in the Earth and Mercury magnetospheres following pressure jumps in Solar wind.

## 1. Introduction

Field-aligned currents (FACs) play an important role in the magnetosphere-ionosphere coupling by transporting the electromagnetic energy to the polar ionosphere. They are classified into three systems with Region-1 being in average the largest. Good correlation of its intensity with Solar Wind parameters and its collocation with precipitating magnetosheath particles strongly suggests that Region-1 FAC is generated by the SW interaction with magnetosphere. The ionospheric conditions are another important factor for FAC formation because the current closure is possible only due to finite Pedersen conductance. In winter hemisphere where conductance is lower due to less sunlight large-scale FAC systems are weaker or even absent (Ohtani *et al* 2000) in comparison to summer hemisphere.

It was proposed as early as in (Eastman 1976) that a source of dayside FACs is in the low-latitude boundary layer (LLBL) where transfer of plasma, momentum and energy from the magnetosheath to the magnetosphere takes place. A basic scenario in conditions of quiet SW and without effects of interplanetary magnetic field (IMF) is summarized, for example, in (Cowley 2000). Viscous interaction due to particle scattering and waves transports SW momentum across the magnetopause. Thus, plasma in a thin boundary layer on each flank moves antisunward and stretches frozen magnetic field lines. On the inner side of the layer tangential drag and stress are mapped to the ionosphere along closed field lines establishing convection pattern. Due to highly different conditions and collisions the stress is loaded on ionosphere generating electric field and net cross-polar current.

This dawn-dusk current, corresponding to Region-1 FAC, drags ions antisunward over poles and decelerates tangential plasma motion in the boundary layer far from the Earth.

A number of Global MHD numerical codes were developed to study SW-magnetosphere-ionosphere coupling (for example, Ogino 1986, Fedder and Lyon 1987, Raeder *et al* 1996, Ridley *et al* 2001). Most of them agree that ionospheric conductance regulates magnetospheric circulation and its global state through intensity of Region-1 currents. For quiet conditions at small or zero IMF tracing the current streamlines points out that near magnetopause regions, partly at low and partly at high latitudes, are a source of Region-1 FACs (Tanaka 1995, Janhunen and Koskinen 1997, Siscoe *et al* 2000). Events of sudden pressure jumps in SW generating traveling vortices and associated FACs were a topic of observational case studies (Moretto *et al* 2000, Boudouridis *et al* 2008). In numerical investigation aimed at elucidating the exact mechanism of FAC generation by pressure jump (Keller *et al* 2002) it is reported that the largest by far response is Region-1 system originating in low-latitudes well inside magnetosphere immediately adjacent to the boundary layer. Driver regions first appear at the dayside and move tailward along magnetopause flanks, inducing corresponding shift of FAC sectors in the ionosphere. It was concluded that conversion between compressional and shear Alfven waves is the best candidate for mechanism of FAC generation.

The main features of FACs are well established in the polar regions - sunward component of magnetic field perturbation in the Northern hemisphere (antisunward in the Southern), upward accelerated bursts of electrons extracted from the ionosphere at the downward current sectors (dawnside for Region-1) and downward accelerated electrons inducing auroral emissions at the upward sectors. As predicted by models and numerical simulations (Siscoe *et al* 2002), strong Region-1 current reduces main component of dipole field inside dayside magnetosphere. Statistical analysis of GOES-10 data taking into account conditions of saturated polar cap potential presented in (Borovsky *et al* 2009) seems to confirm this effect.

However, so far nothing is known about FAC related features in a driver region. To the authors' knowledge suitable spacecraft observations were not analyzed in that context, and this is very difficult to do with data since sparse measurements at variable conditions have to be compared against different ionospheric conditions and FAC intensity. While only *in sutu* observations can unequivocally link ionospheric FACs with faraway boundary layers where SW intermixes with dipole magnetic field, in this work we present exactly such evidence obtained in laboratory experiment. As described in the previous paper by the authors (Shaikhislamov *et al* 2009), laser-produced plasma interacting with magnetic dipole, besides forming a well defined dayside magnetosphere, generates intense field aligned current system. Detailed measurements of total value and local current density, of magnetic field at the poles and in the equatorial magnetopause, and particular features of electron motion in the current channels revealed its similarity to the Region-1 system in the Earth magnetosphere. Such currents were found to exist only if they can closure via conductive cover of the dipole. Comparison of conductive and dielectric cases revealed specific magnetic features produced by FAC and their connection with electric potential generated in the equatorial part of magnetopause.

The next natural step following the cited above experiments is to look for a specific magnetic feature associated with FACs in the low-latitude boundary layer, if those currents are connected to it. In fact, this is a tangential component of magnetic field produced by sheared plasma motion. Magnetic measurements with conductive and dielectric dipole covers described in this paper reveal that such a component is indeed present. It has spatial structure across magnetopause with maximum inside magnetosphere and immediately adjacent to boundary layer. Current associated with this tangential component of magnetic field flows upward on the inner side of LLBL at the dawn sector and downward at the dusk sector, corresponding to Region-1 FAC direction.

It should be stated that, while major portion of literature on FACs is concerned with IMF orientation, induced reconnection, flux transfer events and polar-cap potential saturation, in the present work there is no background magnetic field modeling the IMF. Investigation concerns only the dayside magnetosphere as no tail forms during a flow time span of laser-produced plasma. Such conditions are most relevant to FAC generation following dynamic pressure jumps in SW. Laboratory simulation of planetary magnetospheres has a number of well-known limitations such as absence of developed bowshock and relatively large kinetic scales (Schindler 1969). Despite of them MHD magnetospheric generator driving currents along magnetic field lines of dipole seems to operate on the same physical principals in laboratory and space. Thus, experiment described below gives independent confirmation of the basic magnetospheric generator scenario and rather new data to be compared with analytical and numerical models. The paper consists of two sections on experimental set up including problem formulation and experimental results, followed by conclusions.

## 2. Problem formulation and experimental set up

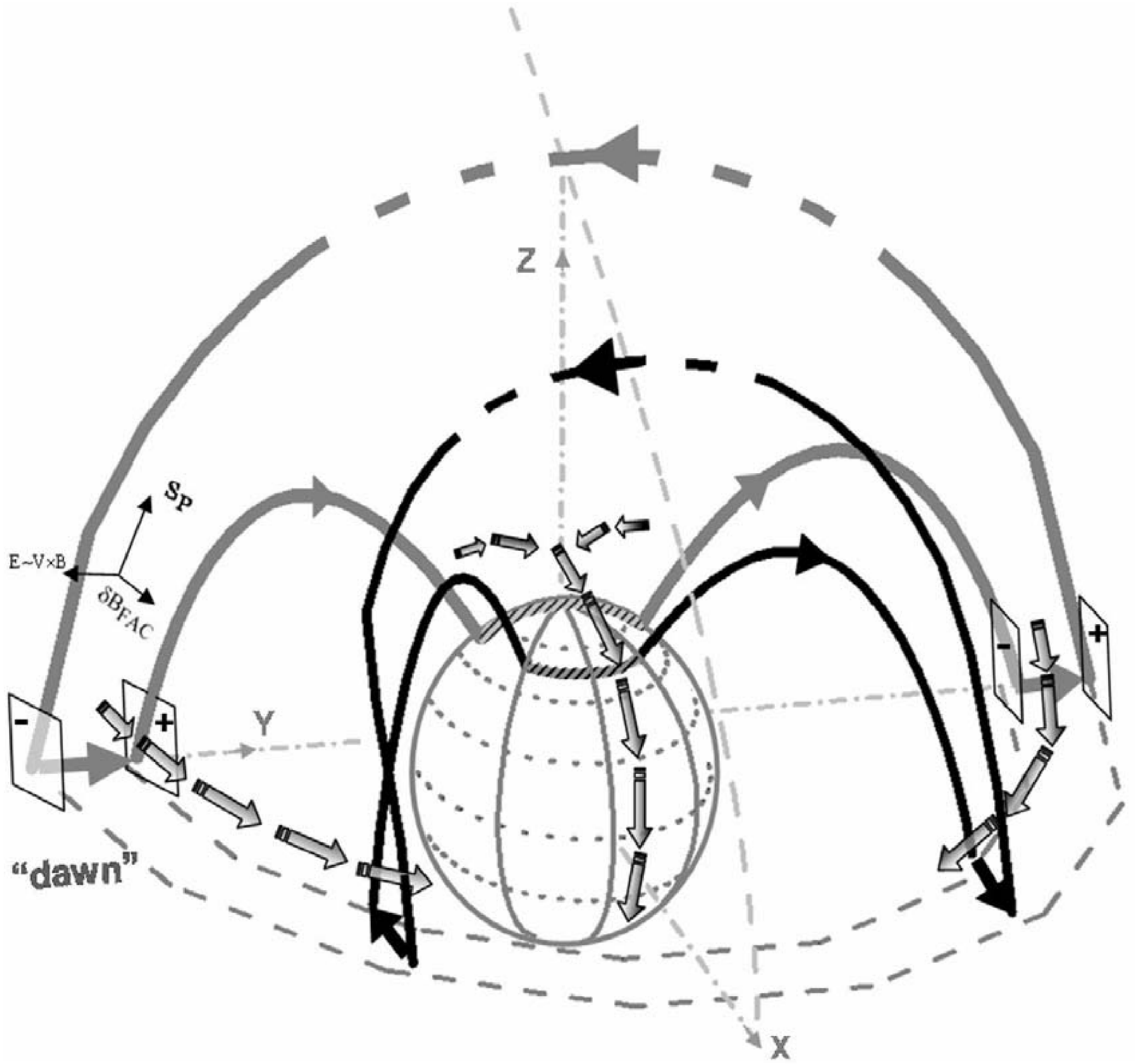

**Figure 1.** A scheme of magnetospheric generator.

Figure 1 demonstrates schematically magnetospheric generator driving Region-1 FAC in the Northern hemisphere. GSM coordinate frame is used throughout the paper as well as a spherical one based on it with athimuthal angle counted anticlockwise from the X-axis. By thick lines two current loops are drawn, in the terminator plane (gray) and topologically similar loop at the dayside (black). Both loops closure through the ionosphere. The return current outside magnetosphere isn't established for certain and some part might flow via high-latitude cusp. On both flanks of the low-latitude layer plasma moves along the boundary across field and with shear along magnetic field. This generates electric field and azimuthal component of magnetic field $\delta B_{FAC}$ such that the Pointing vector $S_p$ is directed upward as shown at the left of the picture. Sequential arrows show that for anticlockwise angle $\varphi$ azimuthal magnetic field $B_\varphi$ produced by tangential drag and sheared stress is positive at the dawnside and negative at the duskside. Note that currents associated with such $B_\varphi$ should flow upward at the inner side of boundary layer at dawn and downward at dusk sector. Sequential arrows in the meridian plane depict magnetic field produced by FAC loops inside magnetosphere – sunward over the North pole and opposite to the dipole field at the equator.

According to the picture above the aim of the present work was to measure azimuthal magnetic field in relation to FACs. Experiment has been carried out at KI-1 space simulation Facility, which includes chamber 5 m in length and 1.2 m in diameter with operating base pressure $10^{-6}$ Torr. Two $CO_2$ beams of 70 ns duration and 150 J of energy each were focused and overlapped into a spot 1 cm in diameter on surface of a solid target made of perlon ($C_6H_{11}ON$). Laser-produced plasma consisted mostly of $H^+$ and $C^{4+}$ ions approximately in equal parts and expanded inertially in a cone ≈1 sr with an average velocity $V_o \approx 1.5 \cdot 10^7$ cm/sec. A total kinetic energy and a number of ions in the plasma was ≈40 J and $\approx 5 \cdot 10^{17}$ respectively. At the axis of plasma expansion at a distance of 60 cm magnetic dipole was placed. A value of magnetic moment was $\mu = 1.15 \cdot 10^6$ $G \cdot cm^3$; a fall off time $\sim 10^{-3}$ sec while a typical time of interaction duration $\sim 10^{-5}$ sec. Magnetic moment was always oriented perpendicular to the interaction axis either to North or South. The dipole has a stainless cover shield in form of a sphere 8 cm in radius. At a time of about t=2 μs after laser irradiation of the target plasma reaches a deceleration region and at a distance of about $R_m$=15 cm a well defined magnetopause is formed. Due to specifically adjusted pulse and tail generation mode of laser oscillator, main plasma flow is followed by a second pressure jump. It slightly compresses magnetosphere and causes pronounced response in field-aligned currents. Such conditions resemble a pressure jump in the SW. Kinetic pressure of laser-produced plasma measured by electric probe will be given below.

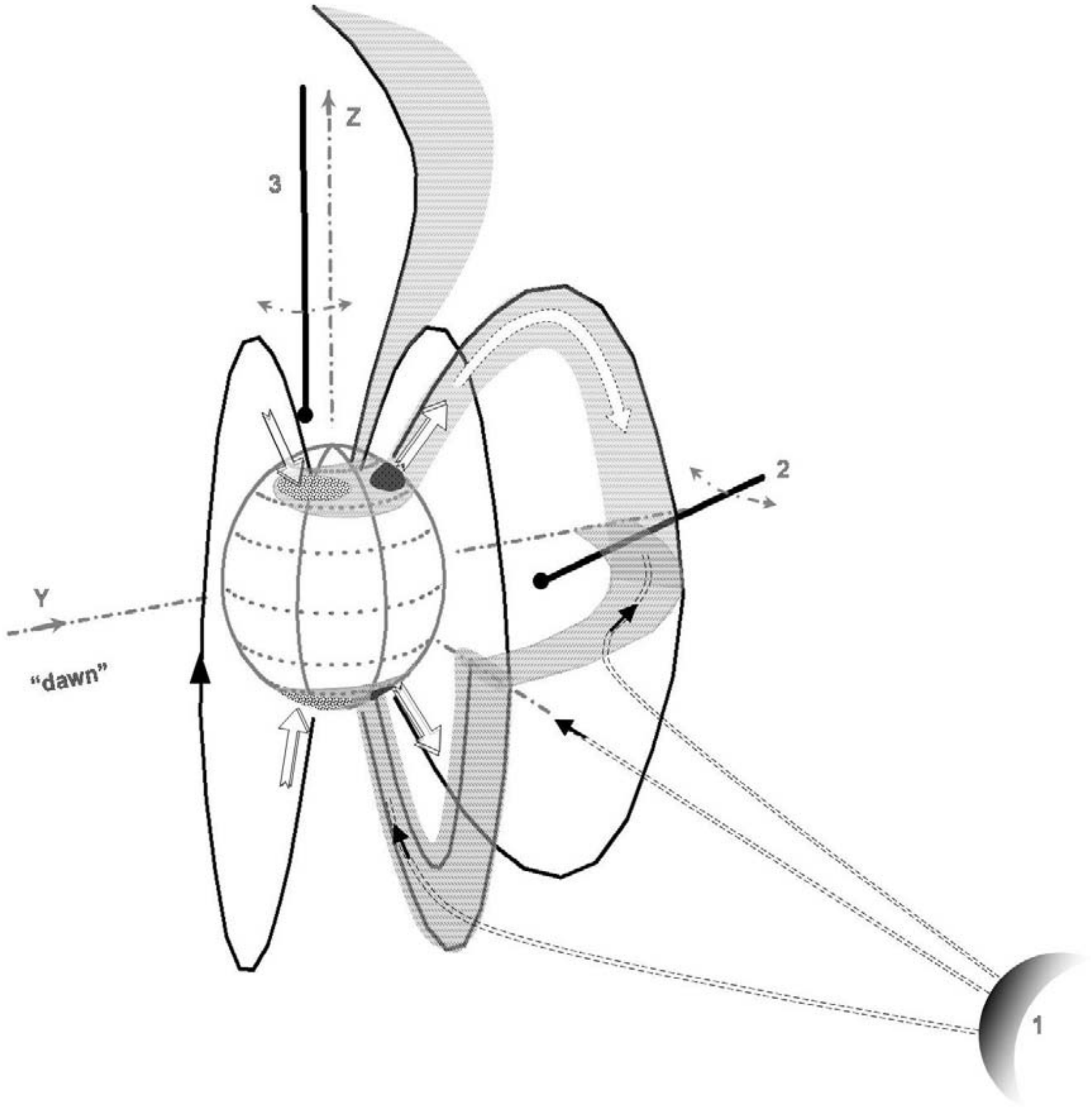

**Figure 2.** Experimental set up. 1 – laser-produced plasma; 2 – LLBL magnetic probe; 3 – Polar magnetic probe. Also schematically are shown magnetic field lines (black), plasma streamlines (double lines), plasma-field boundary layers and polar ovals of plasma precipitation (checked areas), field aligned currents (white arrows) and spots of specific luminosity induced by them at the dipole cover.

As shown in figure 2 and described in detail in (Shaikhislamov *et al* 2009) in such experimental set up strong FAC system develops over poles. Its density was measured directly by Rogovski coil, while a total value by a special double plate construction. Judging from the total value (only 2−4 times smaller than total Chapman-Ferraro current) and very high conductance of metal cover playing the role of ionosphere the FAC system could be in saturation. As stated in the introduction and as will be shown in the present experiment, magnetic perturbation over poles and near equator at the dayside well corresponds to the Region-1 system. For laboratory experiments the precipitation oval is rather wide ($\theta$=40−70$^o$ latitudes) and stretches over most of dayside sector. These low latitudes are probably the result of relatively large dipole size in comparison to magnetosphere. We note that such situation is typical of Hermean magnetosphere for which the influence of FACs on the structure and dynamics was formulated as one of the main problems in Mercury investigations (Baumjohann *et al* 2006). The channels of FAC flow into and out of the dipole cover are inside of precipitation oval and are marked by specific luminosity induced by current. They map into magnetosheath and adjacent parts of magnetosphere.

It was also found that a dielectric film laid over metal casing of the dipole suppresses FACs. This fact allowed extracting the input of FACs into magnetospheric field. Unlike the spacecraft measurements, magnetic probes in laboratory experiments record the change of field due to formation of magnetosphere and all other processes combined. Magnetosphere compression and Chapman-Ferraro currents induce magnetic field perturbation to be directed sunward over the North pole (as FACs input) and strengthens dipole field at the equator (opposite to FACs input). Because experimental magnetosphere forms only at the dayside and doesn't extend in the tail, all components of magnetic field including azimuthal one are present off meridian and equator planes. This $B_\varphi$ component generated by magnetopause currents has the same symmetry as the FAC related sheared stress field. However, it should have essentially different spatial and temporal features. Comparison of dielectric and conductive dipole covers makes it possible to deduce FAC related $B_\varphi$ which is the aim of the present work.

To do this, LLBL magnetic probe was placed above the equator plane (Z=4 cm) and off the meridian plane at the duskside at an azimuth of $\varphi \approx 15-20^{O}$. While for such angles it is more appropriate to use terms of afternoon and postnoon, to keep the meaning straight we'll call it dusk and dawnside. Starting from position well inside magnetosphere the probe could cross boundary layer and then beyond in undisturbed plasma flow, keeping the azimuth approximately the same. The probe could be also pulled out across meridian plane to the dawnside and make the boundary crossing. The other way to measure at the dawnside was to invert dipole moment to the North and take into account that all signals are inverted as well. The Polar probe was placed almost directly above dipole axis at height Z=12 cm and could be shifted by about 5 cm along terminator line. The current that flows in artificial ionosphere through a noon-meridian cut, that is total cross-polar current, was measured by the same way as described in (Shaikhislamov *et al* 2009). Two thin copper plates were laid on dusk and dawn sides of the dipole pole. They were electrically insulated from the dipole casing and separated from each other along the interaction axis. A shortcutting shunt combined with Rogovski coil measured total current flowing between plates.

## 3. Experimental results

In figure 3 typical LLBL probe oscilloscope signals are shown well inside magnetosphere (second panel), near boundary layer and outside magnetosphere (forth panel). On each panel three components of magnetic field perturbation are given. One can see that $\delta B_r$ component is everywhere positive. This is because plasma tends to flatten dipole field lines near equatorial plane. The main $\delta B_\theta$ component reverses sign because dipole field is compressed inside magnetosphere and is totally expelled outside. There is also the azimuthal component. Note that it is of the same sign as shown in figure 1 at the duskside and that in the middle panel it is about three times larger than in others. Another noticeable feature in signals is two maxima around 3 and 6 μs. For the last one the second pressure jump in the laser-produced plasma is responsible as can be deduced from the upper panel, where plasma kinetic pressure is given.

In figure 4 two panels show three component signals at the dawnside ($\varphi \approx -20^{O}$) and the duskside ($\varphi \approx +17^{O}$) at approximately the same distance from the dipole center $R = 16$ cm. The $\delta B_r$ and $\delta B_\theta$ components are of the same sign and, except some shot to shot variability, rather similar. Azimuthal field $\delta B_\varphi$ on the other hand shows unmistakable reversal. Suchlike measurements with dipole moment oriented backwards to the North show the same picture but with *all* components reversed. Thus, azimuthal field $\delta B_\varphi$ has quadruple structure and changes sign at crossing either meridian or equator plane.

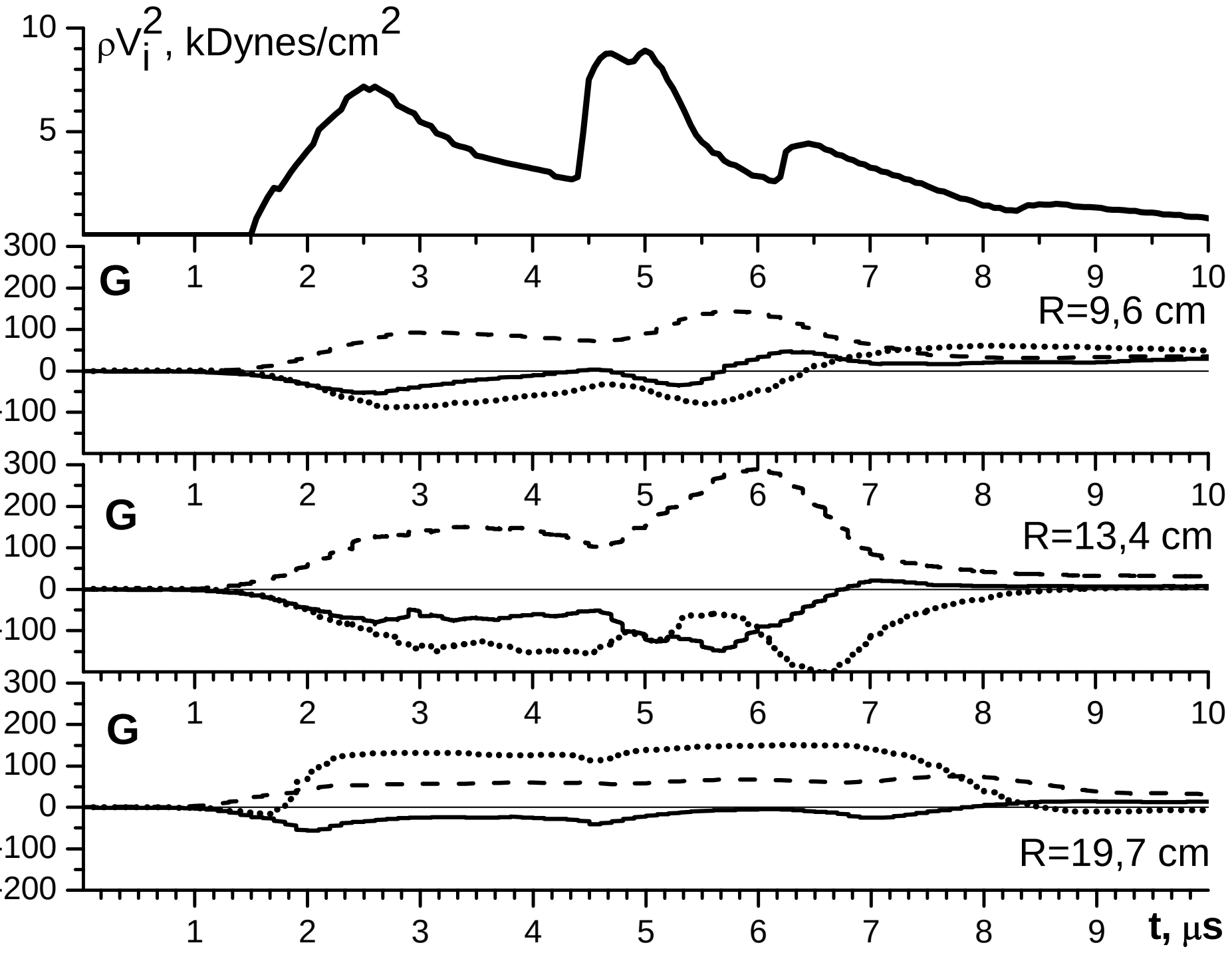

**Figure 3.** Three component oscilloscope signals of LLBL probe at different locations at the dusk sector for conductive dipole cover. Solid line - $\delta B_\varphi$, dashed - $\delta B_r$, dotted - $\delta B_\theta$. Upper panel shows dynamics of plasma kinetic pressure measured by electric probe at a distance of R=10 cm in the absence of dipole field.

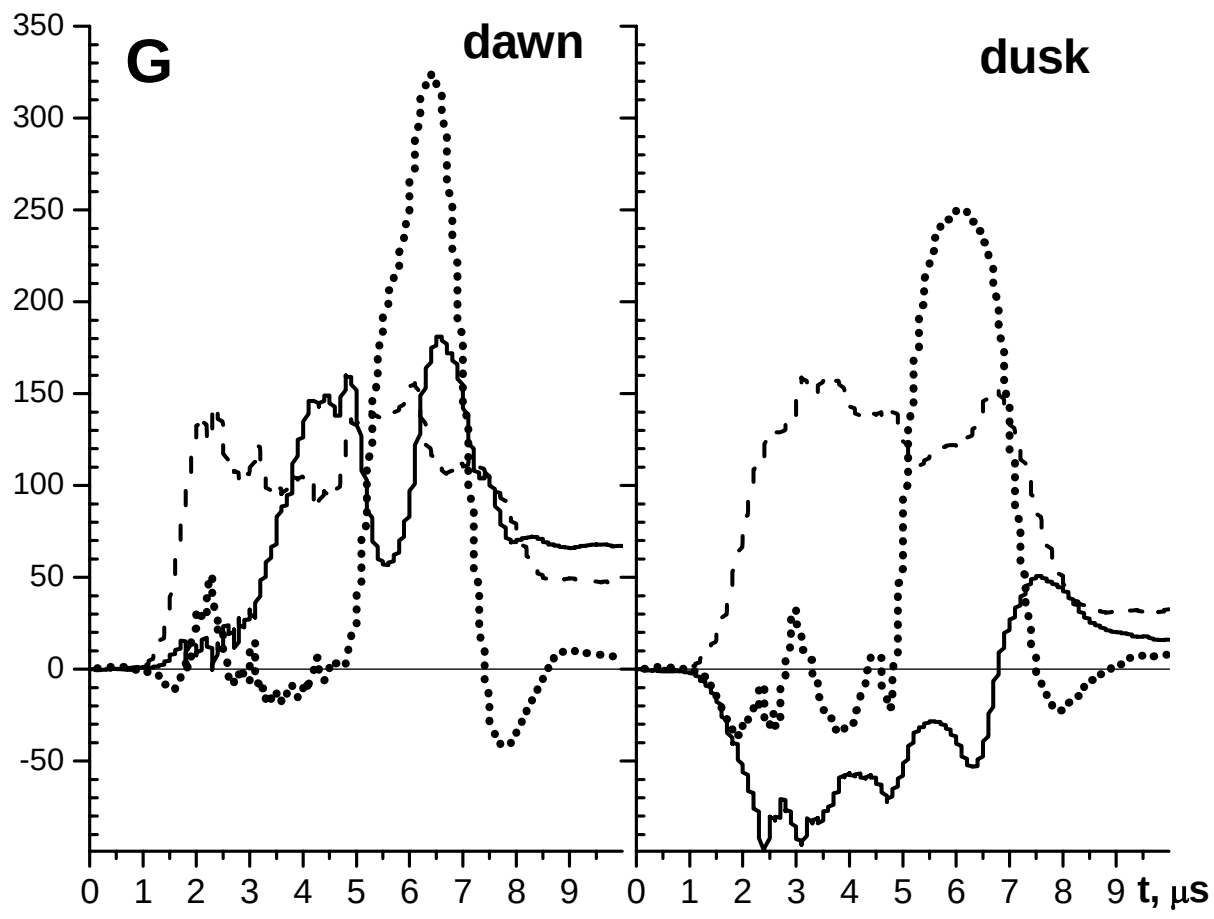

**Figure 4.** Three component oscilloscope signals of LLBL probe at opposite sides of the meridian plane. Solid line - $\delta B_\varphi$, dashed - $\delta B_r$, dotted - $\delta B_\theta$.

The essence of the difference between dielectric and conductive dipole covers is revealed in spatial profiles plotted from signals at different positions. These profiles at the duskside are given in figure 5 for a number of sequential times. From the $\delta B_\theta$ component (upper panels) one can see how the plasma forms magnetosphere and how the second pressure jump pushes it slightly closer to dipole. The boundary layer (from signal minimum to maximum) is about 3−4 cm wide. Field compression inside magnetosphere is substantially reduced for the conductive case. The second pressure jump makes the field even smaller than initial dipole one. This is interpreted as the effect of Region-1 FAC discussed above. The azimuthal component $\delta B_\varphi$ is presented in the lower panels. For the dielectric case it shows oscillations and general increase by amplitude towards dipole. For the conductive case however, there is a much more pronounced structure – a compact minimum inside magnetosphere but immediately adjacent to the boundary layer. It is formed solely by the second pressure jump.

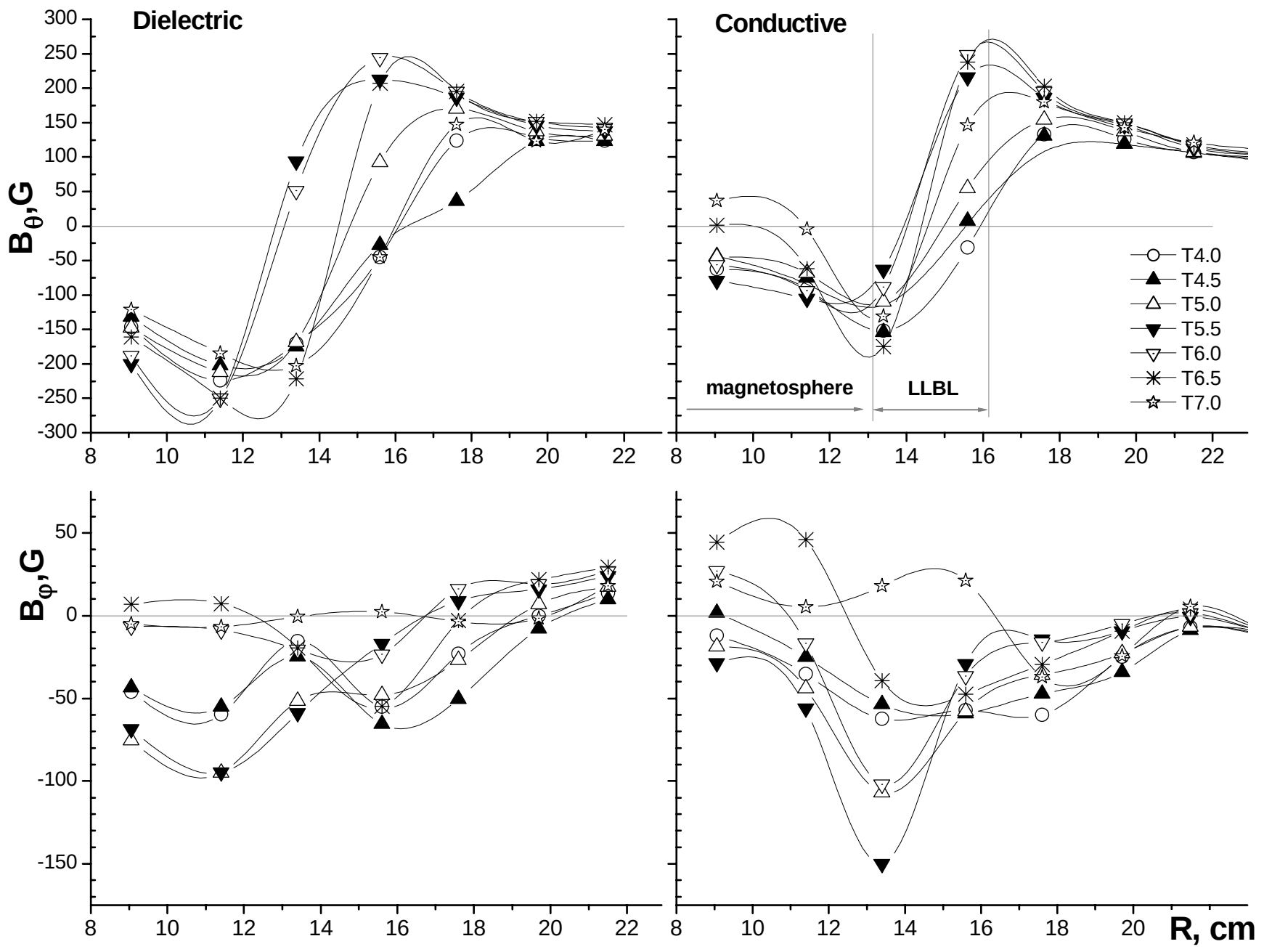

**Figure 5.** Spatial profiles of $\delta B_\theta$ and $\delta B_\varphi$ components at the duskside at different times for cases of dielectric and conductive dipole cover. On the upper right panel there are indicated magnetosphere region and boundary layer.

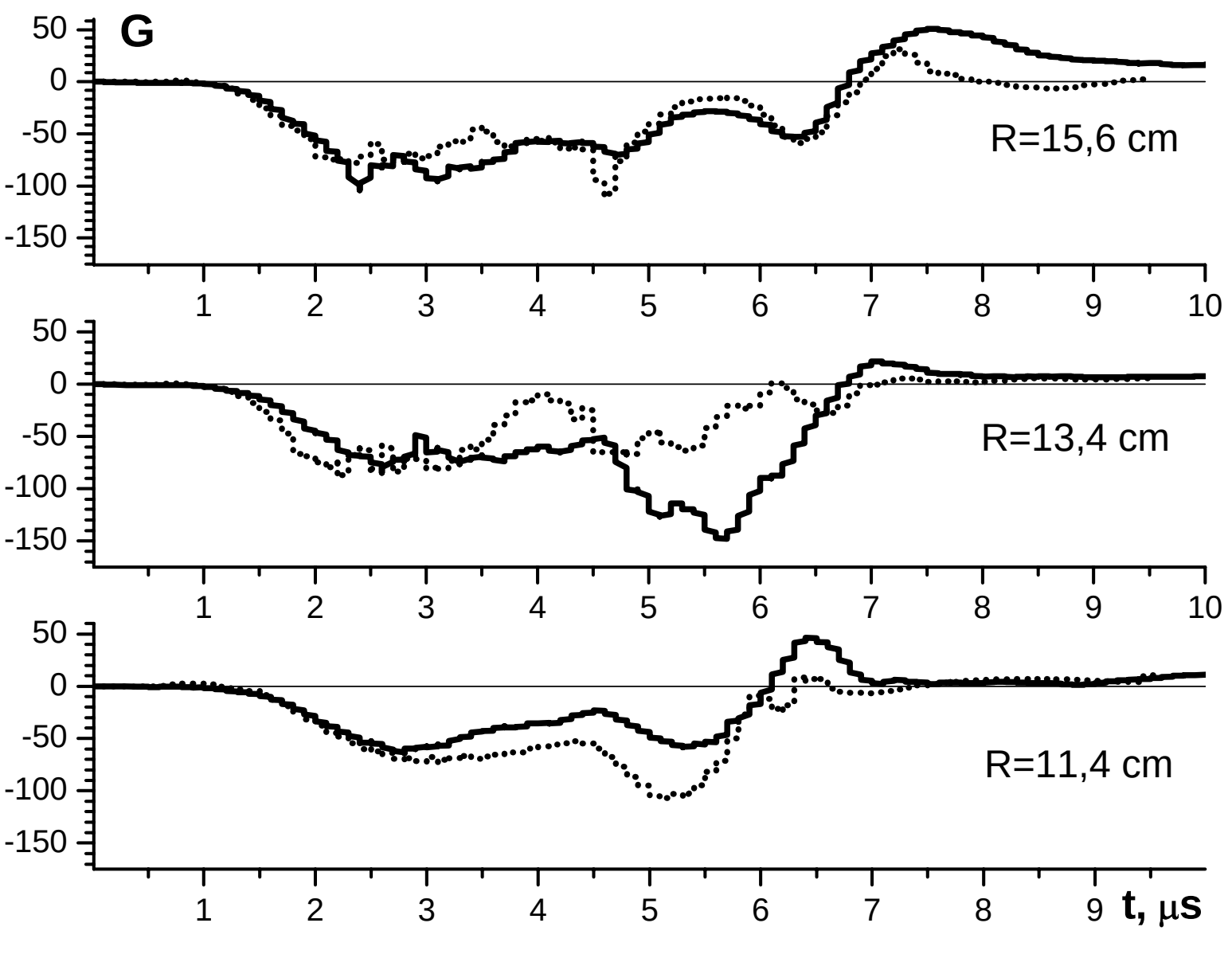

**Figure 6.** Oscilloscope signals of $\delta B_\varphi$ component at three different locations at the duskside for the dielectric (dotted curves) and conductive (solid) cases.

This effect could be seen in more detail at oscilloscope signals. As shown in figure 6, $\delta B_\varphi$ is approximately the same within boundary layer, markedly larger by amplitude for conductive case on the inner side of the layer (middle panel) and smaller inside magnetosphere. Approximately the same tendencies demonstrated in figures 5 and 6 were observed at the dawnside.

The Polar probe, being closer to intense FACs, showed more pronounced difference between dielectric and conductive cases. This is demonstrated in figure 7 for the sunward component of magnetic field $\delta B_X$. It is by two-three times larger when FACs are present and lasts much longer. The $\delta B_y$ component was also substantially larger, while $\delta B_Z$ changed sign. In dielectric case $\delta B_Z$ is produced by diamagnetic effect of plasma flowing along field lines into the pole. Thus, $\delta B_Z$ is opposite to dipole field. In conductive case the part of FAC loop that stretches towards dayside magnetopause produces negative $\delta B_Z$ which is of the same sign as dipole field over the poles. In the first approximation the FAC input into magnetospheric field could be obtained by subtraction of dielectric measurements from conductive ones. For the Polar probe these difference signals for all components are shown in figure 8. The left panel corresponds to the North pole, while the right – to the South obtained by inverting dipole moment.

There also is present another key experimental fact – the total FAC intensity measured between dawn and dusk plates, or total cross-polar current, as described in the previous section. Moment inversion caused practically exact reversal of the total FAC signal. As can be seen from figure 8 current starts at about the time when plasma arrives at poles, but the really large FAC up to 2 kA is generated by the second pressure jump.

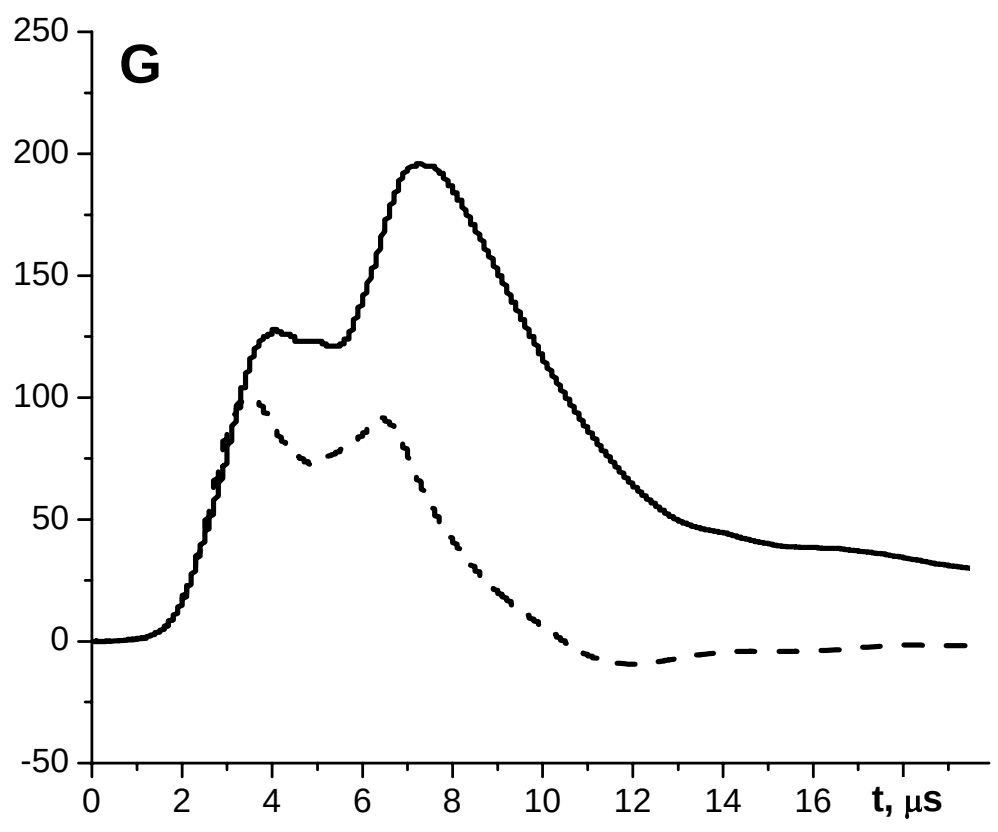

**Figure 7.** Oscilloscope signals of $\delta B_X$ component above the North pole for the dielectric (dashed curve) and conductive (solid) cases.

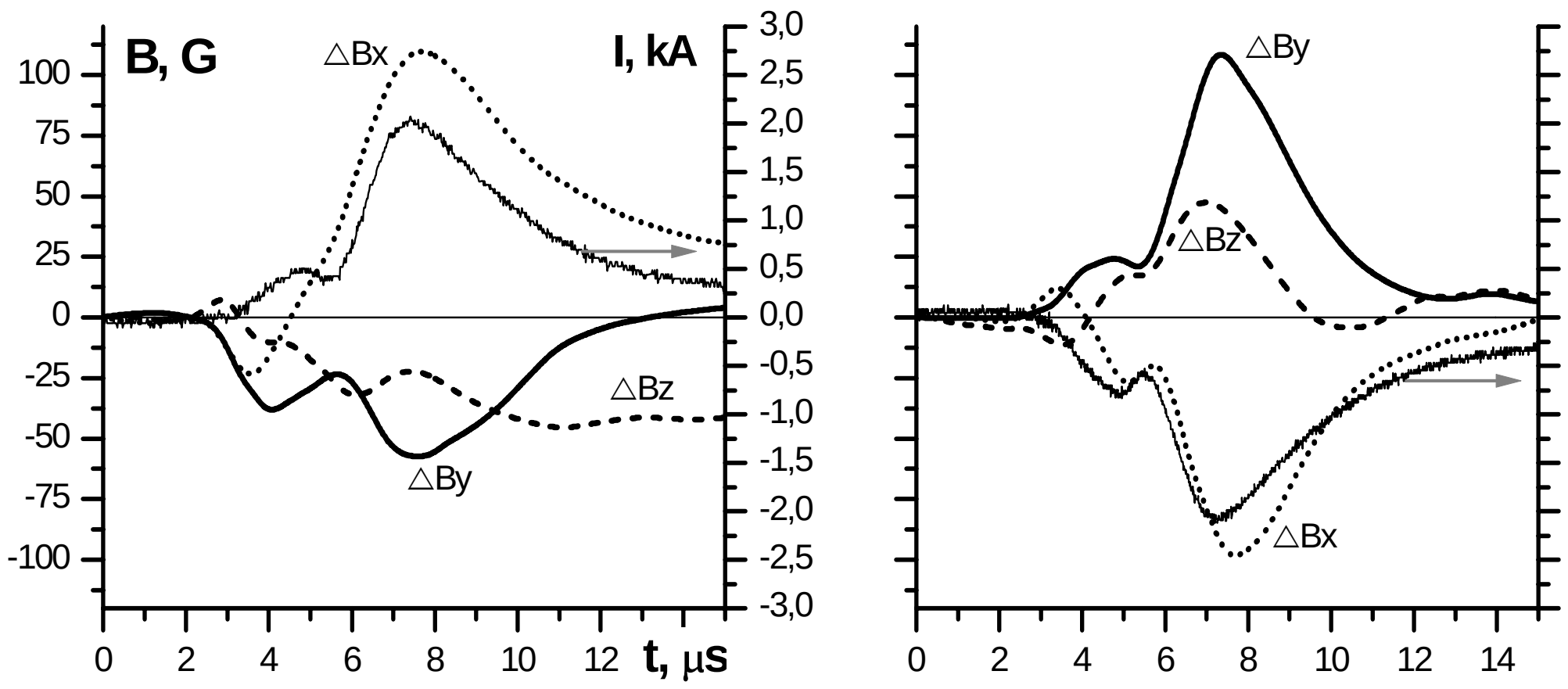

**Figure 8.** Total current measured between dawn-dusk plates (jugged line) over the North pole (left panel) and in case of inverted dipole moment (right). Also shown are difference signals obtained from Polar probe measurements. Solid line - $\Delta B_y$, dashed - $\Delta B_Z$, dotted - $\Delta B_X$.

For the inverted dipole moment magnetic signals also show reversal though, due to experimental conditions, they are not perfectly anti-symmetrical. One can see that difference signals follow more or less closely dynamics of the current. The direction of field vector corresponds to arrows shown in figure 1 above the pole. Measurements along the terminator line (along Y-axis) didn't show any specific structures in distribution of $\delta B_X$ and $\delta B_Z$ components because the probe stayed behind FAC channels and didn't cross them. However, the $\delta B_Y$ component passed zero and changed sign near noon-meridian plane as it should.

## 4. Discussion and conclusions

By taking North and South current loops to have size about 10 cm and total current 2 kA in each, magnetic field produced by FACs in equatorial plane could be estimated as 150−200 G. This is close to the field depression or the difference between dielectric and conductive cases observed in $\delta B_\theta$ component inside magnetosphere (figure5, upper panels). The increase by about 100−150 G of sunward component $\delta B_X$ over the poles induced by FACs (figure 8) also agrees within a factor of two with measured cross-polar current. Spatial structure of azimuthal field (figure5, lower right panel) suggests current which flows downwards on the inner side of low latitude boundary layer (for dusk sector). Total downward current can be estimated as

$$I_{LLBL} \approx c\frac{\Delta B_\varphi}{4\pi}\cdot\pi\frac{R}{2} \approx 1.5-2.5 \text{ kA}$$

where $\Delta B_\varphi$ is maximum value (100−150 G), $\pi R/2$ - length of dayside half arc where current is distributed, R≈13 cm. This estimation is rather close to total FAC measured as cross-polar current. Thus, experiment provided three facts: The azimuthal magnetic field corresponding to expected sheared stress in low latitude boundary layer is indeed generated. Field-aligned current associated with it quantitatively agrees with independently measured total cross-polar current. Third, when "ionosphere" is non-conductive azimuthal field is also different and doesn't have well defined region of local maximums. This suggests that when FAC can't closure over poles magnetic field lines are dragged by plasma as a whole with zero or small stress shear, while with conductive "ionosphere" the current, instead of equalizing stress over field line, flows by the least resistance path. Dynamically, specific increase of azimuthal field occurs before the measured cross-polar current maximum by about 1−2 μs. The lag between cause and consequence is expected and more or less corresponds to time needed for Alfven wave to propagate from equator to poles.

We believe that presented data give first experimental evidence of LLBL origin of Region-1 FAC generator. They confirm the findings of numerical simulations (Keller *et al* 2002) that position of driving region is well inside magnetosphere adjacent to the boundary layer. Due to experimental conditions bowshock doesn't develop upstream of magnetopause and thermal pressure is small. Thus, in our case a driving force behind FAC generation is kinetic rather than thermal pressure gradient. Such a model based on coupling of kinetic pressure variations at magnetopause with Alfven waves was proposed in (Glassmeier and Heppner 1992) with the aim to explain observed transient geomagnetic variation. As was mentioned in introduction, in numerical study of (Keller *et al* 2002) this mechanism was also found to be most relevant for FAC generation by pressure jumps. In the present experiment the shift of magnetopause due to second pressure jump is obviously a compression wave. While no Alfven wave was actually measured in experiment the azimuthal component, being perpendicular to main dipole field and having shear along it, could be interpreted as such. Correlation analysis of magnetic probe signals shows that maximum of azimuthal component moves ahead and synchronously with magnetopause current. This gives independent experimental evidence to the scenario that compressional motion of boundary layer drives shear Alfven waves which carry FACs to the ionosphere.

At this point it is worth while to discuss applicability of laboratory simulation to planetary magnetospheres. At the interaction region the ion density measured by electric probe was $n_i \approx 2.5\cdot10^{12}$ for the main plasma flow and $\approx 8\cdot10^{12}$ cm$^{-3}$ at the second pressure jump, while plasma velocity $V_o \approx 2\cdot10^7$ cm/sec $V_o \approx 10^7$ cm/sec correspondingly. For the lowest estimated electron temperature $T_e \approx 5$ eV magnetic Reynolds number is in excess of 100. Knudsen number calculated by ion-ion and ion-electron collisions also exceeds 100. Thus, collisional parameters of experiment satisfy requirement of physical similarity in a sense of being much larger or much smaller than unity. The other important parameter is ion gyroradius. At position $R_m \approx 13$ cm where generation of azimuthal field occurs magnetic field strength is a sum of dipole field ≈520 G and a part due to compression ≈150 G (figure 5). We note that FAC and azimuthal field generation takes place due to second pressure jump. Taking corresponding velocity, gyroradius calculates as ≈1.5 cm for hydrogen ion and ≈4.5 cm for $C^{4+}$. In average these values are comparable to the width of the azimuthal field structure ≈3 cm but substantially smaller that its lateral dimensions ≈13 cm. In the previous experiments by the authors (Shaikhislamov *et al* 2009) carbon ion gyroradius at the inner edge of the boundary layer varied from 3 to 5.5 cm, while the plasma stand off distance varied from 18 to 28 cm. For all cases essentially the same picture of FAC generation was observed. Moreover, total FAC value in dependence on magnetic moment revealed Chapman-Ferraro scaling $I_{FAC} \sim \mu^{1/3}$. This suggests MHD nature of the FAC generation process, while possible influence of kinetic effects is small.

As was mentioned above total FAC on both poles (≈4 kA) is comparable to total Chapman-Ferraro current (≈9 kA). In future it is planned to find out whether FAC is saturated and to measure its voltage-current characteristic by varying resistance between dawn-dusk plates. It is interesting to see whether saturated FAC alters plasma motion in the low latitude boundary layer. The present experiment has some limitations which are planned to be remedied in future. For example, azimuthal field was measured only near noon-meridian cut though it might extend well to the terminator plane. Laser-produced plasma is impulsive. Plasma flow of a theta-pinch creating stationary magnetosphere combined with laser-produced plasma generating pressure jump is a more suitable set up. Such experiment is aimed to model extreme compression of the Earth magnetosphere by plasma flow of powerful Coronal Mass Ejection (Ponomarenko *et al* 2007, 2008, Zakharov *et al* 2008). We note that CMEs often produce large and saturated polar-cap potential as observed by satellites (Borovsky and Denton 2006), and in numerical simulations of CME impact on the Earth (for example, Ridley *et al* 2006) strong response in Region-1 currents is also a prominent feature. Laboratory experiments are able to provide independent data and new insights into physics of such phenomena.

Acknowledgements: This work was supported by SB RAS Research Program grant 2.3.1.10, Russian Fund for Basic Research grant 09-02-00492 and OFN RAS Research Program 15.